\documentclass[journal]{IEEEtran}

\ifCLASSINFOpdf
\else
   \usepackage[dvips]{graphicx}
\fi

\usepackage[
    hidelinks,
    pdfborder={0 0 0},
    colorlinks=true,
    linkcolor=blue,
    citecolor=blue,
    urlcolor=blue
]{hyperref}

\usepackage{graphicx}
\usepackage{cite}
\usepackage{amsmath,amssymb,amsfonts}
\usepackage{algorithmic}
\usepackage{graphicx}
\usepackage{hyperref}
\usepackage{textcomp}
\usepackage{array}
\usepackage{dsfont}
\usepackage{subcaption}
\usepackage{xcolor}
\DeclareMathOperator*{\argmin}{\arg\!\min}

\begin{document}

\title{Toeplitz-Hermitian ADMM-Net for DoA Estimation}

\author{Youval Klioui

	% <-this % stops a space
	\thanks{This work was funded by the RAISE collaboration framework between
	Eindhoven University of Technology and NXP Semiconductors, including a PPS-supplement from the Dutch Ministry of Economic Affairs and Climate Policy. (Corresponding author: Youval Klioui.)}% <-this % stops a space
\thanks{Youval Klioui is with the Departement of Electrical Engineering, Eindhoven University of Technology, Eindhoven 5612 AP, The Netherlands (email: y.klioui@tue.nl)}% <-this % stops a space
}

\markboth{}
{Shell \MakeLowercase{\textit{et al.}}: Bare Demo of IEEEtran.cls for IEEE Journals}
\maketitle

\begin{abstract}
	This paper presents Toeplitz-Hermitian ADMM-Net (THADMM-Net), a deep neural network obtained by deep unfolding the alternating direction method of multipliers (ADMM) algorithm for solving the least absolute shrinkage thresholding operator problem in the context of direction of arrival estimation. By imposing both a Toeplitz-Hermitian as well as positve semi-definite constraint on the learnable matrices, the total parameter count required per layer is reduced from $\mathbf{N^{2}}$ to approximately $\mathbf{N}$ where $\mathbf{N}$ is the length of the dictionary used in the sparse recovery problem. Numerical simulations show that with a lower parameter count and depth, THADMM-Net outperforms Toeplitz-Lista with respect to the normalized mean-squared error, the detection rate, as well as the root mean-squared error over a signal-to-noise ratio between 0 dB and 35 dB.
\end{abstract}


\IEEEpeerreviewmaketitle
\section{Introduction}
\IEEEPARstart{A}{mong} the different methods known to date for direction of arrival estimation (DoA),   sparsity-based gridded methods such as the least absolute shrinkage thresholding operator (LASSO) \cite{lasso}\cite{candesrobust} offer superior performance compared with subspace-based methods  such as multiple signal classification (MUSIC)\cite{smusic} or estimation of signal parameters through rotational invariance (ESPRIT)\cite{esprit} as they accommodate the single snapshot framework naturally while also not being restricted to uniform linear array topologies (ULA), a topology which is necessary to perform spatial smoothing. LASSO-based methods are furthermore less computationally intensive than both gridless-based methods such as atomic norm minimization (ANM)\cite{anm} and re-weighted atomic norm minimization (RAM) \cite{ram} as well as maximum-likelihood methods (MLE) \cite{mle}, since the former methods require an eigen-decomposition costing $\mathcal{O}(M^{3})$ operations, where $M$ is the array size, per each iteration to perform  a projection onto the set of positive semi-definite matrices (PSD), whereas the latter requires an extensive grid search. 
\par Typical methods for solving LASSO such as the iterative shrinkage thresholding algorithm (ISTA)\cite{ista},  the alternating direction method of multipliers (ADMM)\cite{admm}, and coordinate descent\cite{coordinate_descent} often require multiple iterations to converge \cite{num_iterations}. In order to address this, deep unfolding \cite{unroll} was later on proposed to reduce the number of iterative steps by training a neural  network whose structure is identical to that of the iterative algorithm. The resulting network obtained through this procedure will typically achieve a much better reconstruction  performance for the same number of iterative steps compared with the original algorithm. Although the number of iterations is reduced, the number  of parameters  that need to be learned can still be substantial. For instance, Learned-ISTA (LISTA)\cite{unroll}, a deep unfolded network based on ISTA, requires approximately   $N^{2}+NM$ parameters per layer, where $M$ and $N$ are the size the measurement vector $\mathbf{y}$, which is the same as the array size, and the length  of the dictionary matrix $\mathbf{A}$ in ISTA, respectively.
\par Structured approaches to deep unfolding have been introduced to address this issue. Motivated by the fact that the Gram of $\mathbf{A}$, $\mathbf{A}^{H}\mathbf{A}$, is a Toeplitz  matrix, Topelitz-LISTA (TLISTA) \cite{tlista} imposes a such a structure on the learnable mutual inhibition matrix during the training phase, thereby bringing down the total  count of parameters per layer to $2N-1+NM$, while maintaining a performance close to LISTA. We similarly propose here a structured approach to the deep unfolding of ADMM by imposing both a Hermitian as well as Toepltiz constraint on the learnable matrix, the resulting network termed THADMM-Net whereby the total learnable parameters per layer is reduced from $N^{2}$ for ADMM-Net  down to $N$. The main contributions of this paper can be summarized as follows:
\begin{itemize}
\item We introduce THADMM-Net, a deep-unfolded network based on ADMM that makes use of learnable matrices with a Toeplitz-Hermitian as well as PSD constraint, thereby significantly lowering the memory footprint per layer compared with ADMM-Net.
 \item We examine the performance of the proposed network and show that it manages to outperform TLISTA, THLISTA, LISTA, and ADMM-Net with respect to the NMSE, the detection rate, and the RMSE for an SNR level ranging from 0 db up to 35 dB.
\end{itemize}
\par The next section briefly reviews the ADMM solution for LASSO within the context of DoA estimation and details the architecture of THADMM-Net, 
The third section details the training procedure and provides a performance comparison against TLISTA with respect to the detection rate, the RMSE, and  
the NMSE, and the fourth section concludes this paper.\footnote{The repository for replicating the results reported here can be found at: \url{https://github.com/youvalklioui/thadmmnet }}

\section{Unrolled ADMM FOR LASSO}
\label{sec:format}
\subsection{Signal Model}
We consider a linear array with elements positioned at $d_{1}, d_{2}, \hdots, d_{M}$ where each position is expressed as a multiple of a fraction of the wavelength $d_{m} =  p_{m}\gamma\lambda$, $p_{m} \in \mathbb{N}$ and $0<\gamma<1$, along with the following signal model,
\begin{align}
	\mathbf{y}=\sum_{k=1}^{K}x_{k}^{*}\mathbf{a}(f_{k}^{*})+\mathbf{n}
\end{align}
where $\mathbf{y} \in \mathbb{C}^{M}$ is the single snapshot measurement vector, $\mathbf{n} \sim \mathcal{CN} (0, \sigma^{2}\mathbf{I})$ is complex Gaussian noise,  $x_{k}^{*} \in \mathbb{C}$ is the amplitude of the $k$-th target located at angle $\theta_{k}^{*}$, and $\mathbf{a}(\theta_{k}^{*})$ is the array steering vector,
\begin{align}
	\mathbf{a}(f_{k}^{*})(m)=  e^{j2\pi p_{m}f_{k}^{*}}, \qquad m=1, 2, \hdots M
\end{align}
where $f_{k}^{*}= -\gamma\sin(\theta_{k}^{*})$ is the frequency associated with the angle $\theta_{k}^{*}$. DoA estimation within the LASSO framework can be formulated as a minimization problem of the following objective function,
\begin{align}
	f(\mathbf{x})=\dfrac{1}{2}||\mathbf{y}-\mathbf{A}\mathbf{x}||_{2}^{2} +\tau ||\mathbf{x}||_{1} \label{lasso}
\end{align}
where $\mathbf{x} \in \mathbb{C}^{N}$ is a sparse vector whose support correspond to estimates of the true frequencies $f_{k}^{*}$, $\tau>0$ is a hyperparameter that balances the sparsity of the solution against the consistency with the observed measurement vector, and the dictionary $\mathbf{A} \in \mathbb{C}^{M\times N}$ is defined as
\begin{align}
	\mathbf{A}= \begin{bmatrix} \mathbf{a}(f_{1}),  \mathbf{a}(f_{2}), \hdots, \mathbf{a}(f_{N})	         \end{bmatrix},
\end{align}
where $\{f_{1}, f_{2}, \hdots,  f_{N}\}$ is the frequency grid and is related to the angular grid $\{\theta_{1}, \theta_{2}, \hdots, \theta_{N}\}$ by $f_{n}=-\gamma\sin(\theta_{n})$.

\subsection{ISTA and TLISTA}
The ISTA solution to the problem in \eqref{lasso} is given by the following fixed-point iterative method \cite{ista},
\begin{align}
	\mathbf{x}^{(t+1)}=S_{\kappa_{1}}\big( (\mathbf{I}-\mu \mathbf{A}^{H}\mathbf{A})\mathbf{x}^{(t)}+\mu\mathbf{A}^{H}\mathbf{y}\big), \label{ista}
\end{align}
where $\mu=1/\sigma_{max}(\mathbf{A})^{2}$, $\sigma_{max}(\mathbf{A})$ is the largest singular value of $\mathbf{A}$,  $S_{\kappa}$ is the soft-thresholding operator defined as $S_{\kappa}(z)\triangleq e^{j\arg(z)}\max(|z|-\kappa)$, and $\kappa_{1}=\mu \tau$ is the threshold level. LISTA \cite{unroll} effectively replaces the iterative procedure in \eqref{ista} with a neural network whose input-output relationship is given by,
\begin{align}
\mathbf{x}^{(t+1)}=S_{\beta^{(t)}}\big( \mathbf{W_{1}}^{(t)}\mathbf{x}^{(t)}+\mathbf{W_{2}}^{(t)}\mathbf{y}\big), \label{lista}
\end{align}
where $\mathbf{W_{1}}^{(t)} \in \mathbb{C}^{N\times N}$ and $\mathbf{W_{2}}^{(t)}\in \mathbf{C}^{N \times M}$ are learnable matrices, and $\beta^{(t)}>0$ is a learnable threshold. Thus LISTA requires $N^{2}+MN+1$ parameter per layer. TLISTA \cite{tlista} \cite{tlista1} exploits the fact that as long as the frequencies $f_{1}, f_{2}, \hdots, f_{N}$ are equally spaced, $\mathbf{A}^{H}\mathbf{A}$ will be Toeplitz, and so will $\mathbf{I}-\mu\mathbf{A}^{H}\mathbf{A}$. Thus the input-output relationship at the $t$-th layer  of TLISTA is identical to \eqref{lista} with the exception that now $\mathbf{W_{1}}^{(t)}$ is constrained to to the set of Toeplitz matrices, thereby bringing down the total number of parameters per layer  to $2N+MN$.

\subsection{ADMM and THADMM-Net}
\begin{figure}[htb]
	\centering
	\includegraphics[height=3.2cm,width=7.5cm]{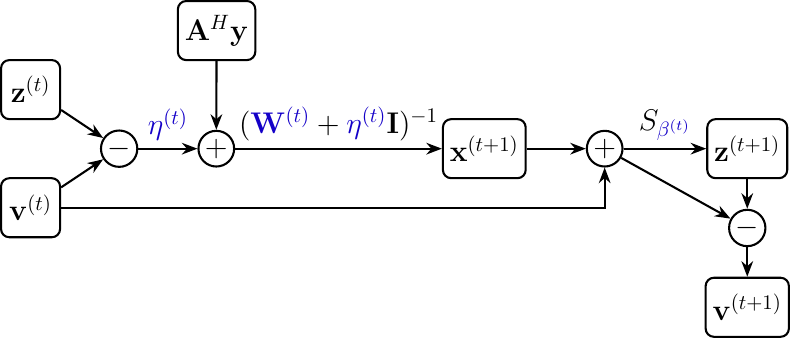}
	\caption{$t$-th layer of THADMM-Net. The learnable parameters are highlighted in blue. }
	\label{fig:schematic}
\end{figure}

The ADMM is another popular iterative algorithm that can be applied to solve the LASSO problem in \eqref{lasso} that tends to exhibit a higher convergence rate when compared with ISTA \cite{num_iterations}. The ADMM solution to the LASSO problem is given by the following iterative procedure\cite{admm},
\begin{align}
	 & \mathbf{x}^{(t+1)}=(\mathbf{A}^{H}\mathbf{A}+\rho\mathbf{I})^{-1}(\mathbf{A}^{H}\mathbf{y}+\rho(\mathbf{z}^{(t)}-\mathbf{v}^{(t)})) \label{admm1} \\
	 & \mathbf{z}^{(t+1)}=S_{\kappa_{2}}(\mathbf{x}^{(t+1)}+\mathbf{u}^{(t)})                                               \label{admm2}\\
	 & \mathbf{v}^{(t+1)}=\mathbf{x}^{(t+1)}+\mathbf{v}^{(t)}-\mathbf{z}^{(t+1)}, \label{admm3}
\end{align}
where $\rho>0$ is a hyperparameter and the threshold level is set to $\kappa_{2}=\rho\tau$. A straightforward deep unfolding of ADMM to yield ADMM-Net would effectively replace the steps in \eqref{admm1} and \eqref{admm2} with,
\begin{align}
	& \mathbf{x}^{(t+1)}=(\mathbf{W}^{(t)}+\rho^{(t)}\mathbf{I})^{\dagger}(\mathbf{A}^{H}\mathbf{y}+\rho^{(t)}(\mathbf{z}^{(t)}-\mathbf{v}^{(t)})) \\
	 & \mathbf{z}^{(t+1)}=S_{\beta ^{(t)}}(\mathbf{x}^{(t+1)}+\mathbf{v}^{(t)}),                                               
\end{align}
where the learnable parameters are  $(\mathbf{W}^{(t)},\rho ^{(t)}, \beta ^{(t)})$ and $(.)^{\dagger}$ denotes the pseudo-inverse operator. The parameter count per layer of ADMM-Net is therefore $N^{2}+2$.  Toeplitz-Hermitian ADMM-Net constrains $\mathbf{W}^{(t)}$ to be a PSD Toeplitz-Hermitian matrix and makes use of the following parametrization,
\begin{align}
	\mathbf{W}^{(t)} \triangleq \mathbf{W}_{TH}^{(t)} + \max(-\lambda_{min}(\mathbf{W}_{TH}^{(t)}), 0)\mathbf{I}  \label{parametrization}
\end{align}
where $\mathbf{W}_{TH}^{(t)} $ is Toeplitz-Hermitian and fully characterized by $N$ parameters, and $\lambda_{min}(\mathbf{W}_{TH}^{(t)})$ is the smallest eigenvalue of $\mathbf{W}_{TH}^{(t)}$. It can be easily verified that the parametrization in \eqref{parametrization} ensures that $\mathbf{W}^{(t)}$ is always a PSD matrix.  The input-output relationship at the $t$-th layer of THADMM-Net for $\mathbf{x}^{(t)}$  is then given by,
\begin{align}
	 & \mathbf{x}^{(t+1)}=(\mathbf{W}_{TH}^{(t)}+\eta^{(t)}\mathbf{I})^{-1}(\mathbf{\mathbf{A}^{H}\mathbf{y}}+\eta^{(t)}(\mathbf{z}^{(t)}-\mathbf{v}^{(t)}))
\end{align}
where $\eta^{(t)} =  \max(-\lambda_{min}(\mathbf{W}_{TH}^{(t)}), 0) + \rho^{(t)}$ and $\rho^{(t)}>0$.   Since  $\mathbf{W}_{TH}^{(t)}$ is PSD, $\mathbf{W}_{TH}^{(t)}+\eta^{(t)}\mathbf{I}$ will be positive definite, allowing us to substitute the pseudo-inverse with the inverse operator,  which can be efficiently evaluated for positive definite Toeplitz matrices \cite {toeplitz2} \cite{toeplitz1}. Fig. \ref{fig:schematic} illustrates the $t$-th layer of THADMM-Net, where the term $\mathbf{A}^{H}\mathbf{y}$ is pre-computed and shared
across all layers.

\section{THADMM-Net Performance Characterization}
\label{sec:pagestyle}

\subsection{Training Setup}
\subsubsection{Datasets}
We set the parameter $\gamma$ that specifies the array geometry to
$1/2$, and we subsample an array of $M=20$ elements  from a ULA with an aperture of $50\lambda/2$ and use a uniform frequency grid $\mathcal{F}=\{f_{1}, f_{2}, \hdots, f_{N}\}$  of $N=256$ bins covering the  range $[-1/2, 1/2)$.
 In order to characterize the performance of THADMM-Net and compare it against TLISTA, a training set consisting of $10^{5}$ measurement vectors and the corresponding ground truth vectors was prepared. A noise-free measurement vector $\mathbf{y}_{c}$ is generated by first randomly selecting the number of targets $K$ between 1 and 8.
  For each target an associated frequency $f^{*}_{k}$ is then randomly generated between $[-1/2,  1/2)$, while making sure that the frequency separation between any two frequencies is always at least $1/M$ \cite{supres}. Afterwards, the amplitude and phase of each target are both drawn from uniform distributions with $|x_{k}^{*}|\sim \mathcal{U}(0,1)$ and $\arg(x_{k}^{*})\sim \mathcal{U}(0,2\pi)$. The corresponding ground truth vector $\mathbf{x}$ is constructed by first finding for each $f^{*}_{k}$ its closest point $f_{k'}$ on the frequency grid, and then ascribing $\mathbf{x}(k')=x_{k}^{*}$. Zero-mean Gaussian noise $\mathbf{n}\sim \mathcal{CN}(0,\sigma^{2}\mathbf{I})$ is finally added to the clean measurement vector $\mathbf{y}_{c}$ so as to obtain an SNR of 15 dB, where
\begin{align}
	\textrm{SNR}_{\textrm{dB}}=10\log\bigg(\dfrac{||\mathbf{y}_{c}||_{2}^{2}}{\sigma^{2}}\bigg).
\end{align}

The same procedure is used to generate a validation set of $2\times 10^{4}$ vectors. Finally, for the performance characterization, we prepare a test dataset consisting of $8\times10^{3}$ vectors by varying the SNR from 0 dB up to 35 dB with a step size of 5 dB and generating $10^{3}$ test vectors at each SNR step, and the generated frequencies of the test set have a minimum separation requirement of $1/(3M)$.

\subsubsection{Network Sizes , Initialization, and Training}
A 15-layer THADMM-Net network is designed for comparison against 30-layer networks of LISTA, ADMM-Net, TLISTA, and THLISTA, where THLISTA  uses a learnable Toeplitz-Hermitian matrix constraint for $\mathbf{W}_{1}^{(t)}$ in \ref{lista}.  All the weight matrices of the networks are initialized using their corresponding values in the iterative algorithms \cite{unroll}. For instance, $\mathbf{W}_{TH}^{(t)}$ is initialized with $\mathbf{A}^{H}\mathbf{A}$ for THADMM-Net. The learnable threshold $\beta^{(t)}$ is initialized with $10^{-1}$ for all networks, and $\rho^{(t)}$ was initialized with $1$ for (TH)ADMM-Net. As a loss function, we use the normalized mean-squared error (NMSE) between the neural network output $\hat{\mathbf{x}}$ and its corresponding ground truth $\mathbf{x}$,
\begin{align}
	\textrm{NMSE}(\mathbf{x},\mathbf{\hat{x}})=\dfrac{||\mathbf{\hat{x}}-\mathbf{x}||_{2}^{2}}{||\mathbf{x}||_{2}^{2}}.
\end{align}
The networks were trained for 30 epochs using the Adam \cite{adam} optimizer with a batch size of 2048 and a learning rate of $10^{-4}$. Table 1 provides a summary for the comparison between THADMM-Net and TLISTA for this setup.

\begin{table}[h]
	\centering
	\caption{Comparison of THADMM-Net and TLISTA for the training set-up}
	\begin{tabular}{|p{1.75cm}|>{\centering\arraybackslash}p{2.7cm}|>{\centering\arraybackslash}p{3cm}|}
		\cline{2-3}
		\multicolumn{1}{c|}{}                & \raisebox{-1pt}{\textbf{THADMM-Net}} & \raisebox{-1pt}{\textbf{TLISTA}}   \\ \hline
		\textbf{\raisebox{-1pt}{Num layers}} & \raisebox{-1pt} {15}                  & \raisebox{-1pt}{30}                \\ \hline
		\textbf{\parbox[c]{2cm}{ MAC}}       & \raisebox{-1pt}{$\mathcal{O}(N^2)$}              & \raisebox{-1pt}{$\mathcal{O}(MN)$} \\ \hline
		\textbf{\parbox[c]{1.0cm}{\vspace{0.1cm}\centering Parameter count \vspace{0.1cm}                                \\ }} & 3870 & 168960 \\ \hline
		\textbf{Initialization}              &
		\parbox[c]{2.6cm}{\centering
		\((\mathbf{W}_{TH}^{(t)}, \beta^{(t)}, \rho^{(t)}) =\)                                                               \\
		\((\mathbf{A}^\text{H} \mathbf{A}, 10^{-1}, 1)\)
		}                                    &
		\parbox[c]{2.8cm}{\centering
		\((\mathbf{W}_1^{(t)}, \mathbf{W}_2^{(t)},\beta^{(t)}) =\)                                                           \\
		$(\mathbf{I} - \mu \mathbf{A}^\text{H} \mathbf{A},$                                                                       \\
		$\mu \mathbf{A}^\text{H}, 10^{-1})$
		}                                                                                                                \\ \hline
	\end{tabular}

	\label{table:comparison}
\end{table}
\begin{figure}[!htb]
	\centering
	\begin{subfigure}{\linewidth}
		\centering
		\includegraphics[width=0.5\linewidth]{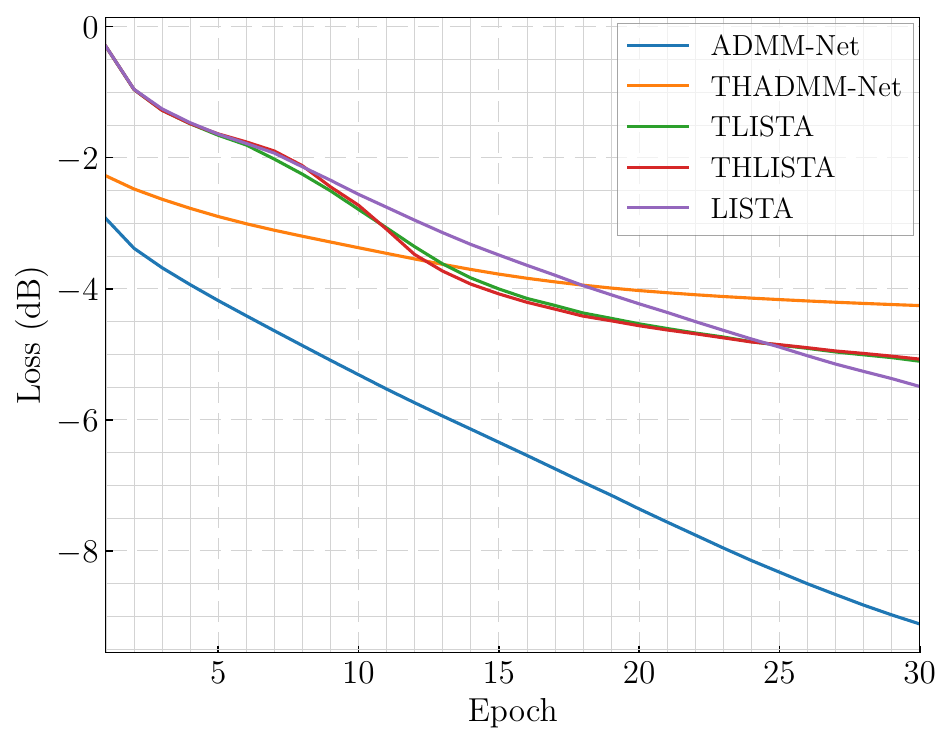}
		\caption{Training loss.}
		\label{training_loss}
	\end{subfigure}
	
	\begin{subfigure}{\linewidth}
		\centering
		\includegraphics[width=0.5\linewidth]{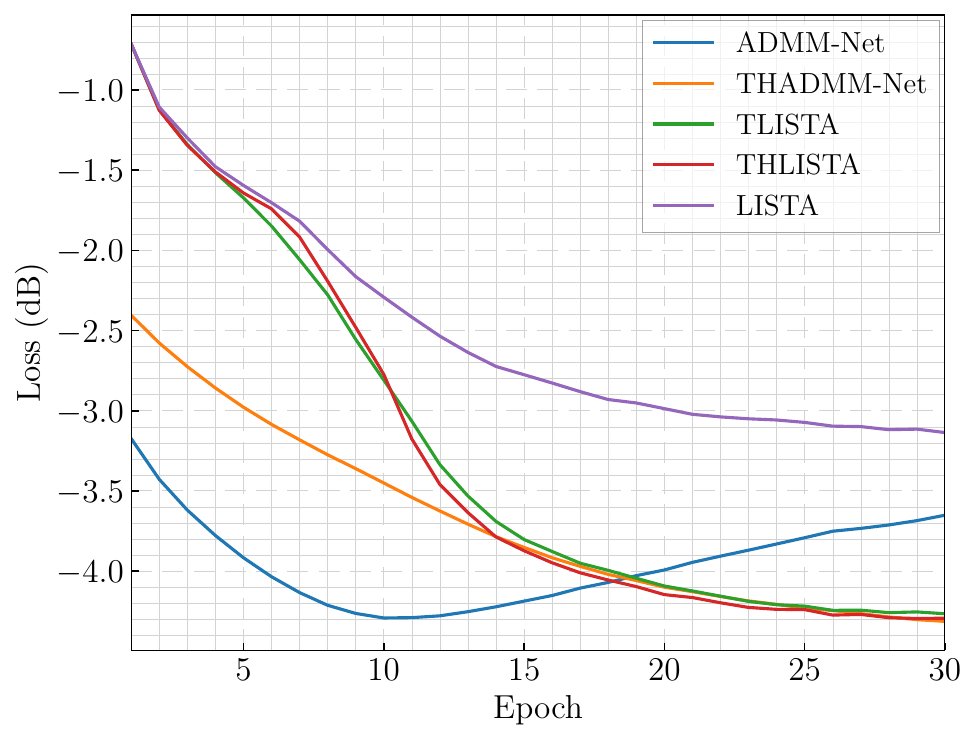}
		\caption{Validation loss.}
		\label{validation_loss}
	\end{subfigure}
	\caption{Training and validation losses for 15-layer THADMM-Net and 30-layer TLISTA, LISTA, THLISTA and ADMM-Net.}
	\label{losses}
\end{figure}
As can be seen from the validation loss curves in Fig. \ref{validation_loss}, THADMM-Net, TLISTA and THLISTA all converge to approximately -4.3 dB after by the 30 epochs mark. We can additionally notice that despite ADMM-Net having a lower parameter count compared with LISTA, it begins to overfit  at a much earlier epoch, at around 10.
\subsection{Peformance Comparaison}
In order to compare the performance of THADMM-Net against TLISTA, we make use 
of the detection rate and the root mean-squared error (RMSE). We first recover an estimate $\mathbf{\hat{x}}$ of the ground truth vector $\mathbf{x}$ associated with a given measurement vector $\mathbf{y}$. Afterwards, we perform peak detection on the amplitude spectrum to obtain a new spectrum $\mathbf{\hat{x}}_{pk}$ containing only the peaks of  $|\mathbf{\hat{x}}|$. Then, for  a   target in the ground-truth spectrum $|\mathbf{x}|$ that is located at a given  index $q$ with respect to the uniform frequency grid, we find the subset of indices $\mathcal{I}_{q}=\{\hat{q}_{1}, \hat{q}_{2}, \hdots, \hat{q}_{L(q)}\}$ from the support of  $|\mathbf{\hat{x}}|$ that all fall within a set distance $\delta_{1}$ from $q$,
\begin{align}
	|q-\hat{q}_{l}| \le \delta_{1}, \qquad l = 1, 2, \ldots, L(q).
\end{align}
Next, among the elements in $\mathcal{I}_{q}$, we retain only the subset of indices $\mathcal{J}_{q} \subseteq \mathcal{I}_{q}$ at which the ratio of the amplitude of
\begin{figure*}[h!]
	\centering{
		\begin{subfigure}{0.3\textwidth}
			\centering
			\includegraphics[width=\linewidth]{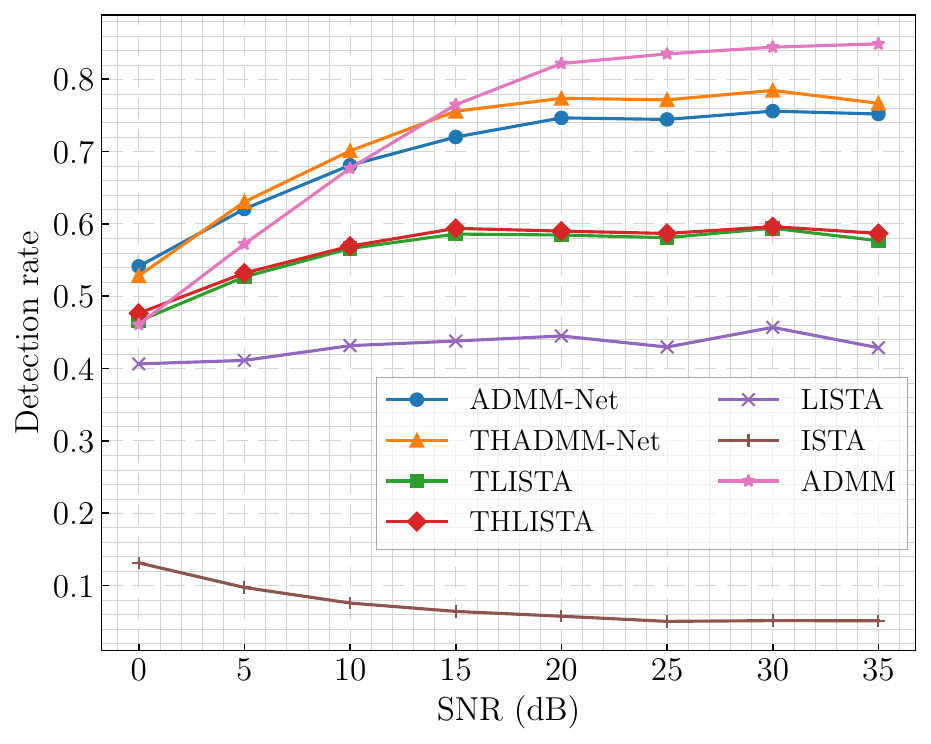}
			\caption{Detection rate.}
			\label{detection_rate}
		\end{subfigure}
		\hfill
		\begin{subfigure}{0.3\textwidth}
			\centering
			\includegraphics[width=\linewidth]{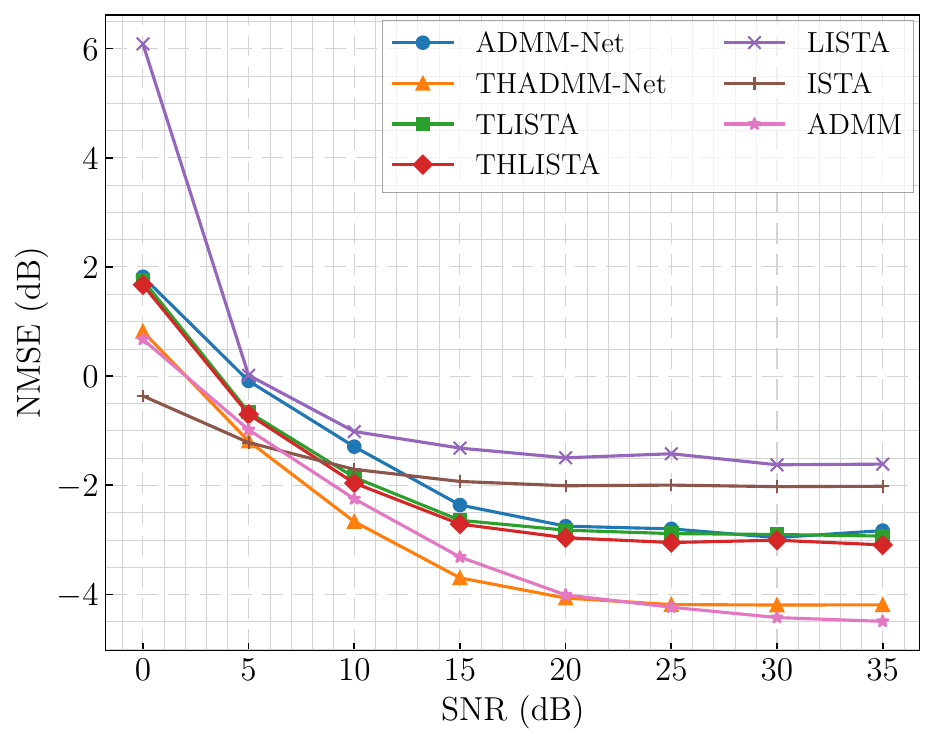}
			\caption{NMSE.}
			\label{nmse}
		\end{subfigure}
		\hfill
		\begin{subfigure}{0.3\textwidth}
			\centering
			\includegraphics[width=\linewidth]{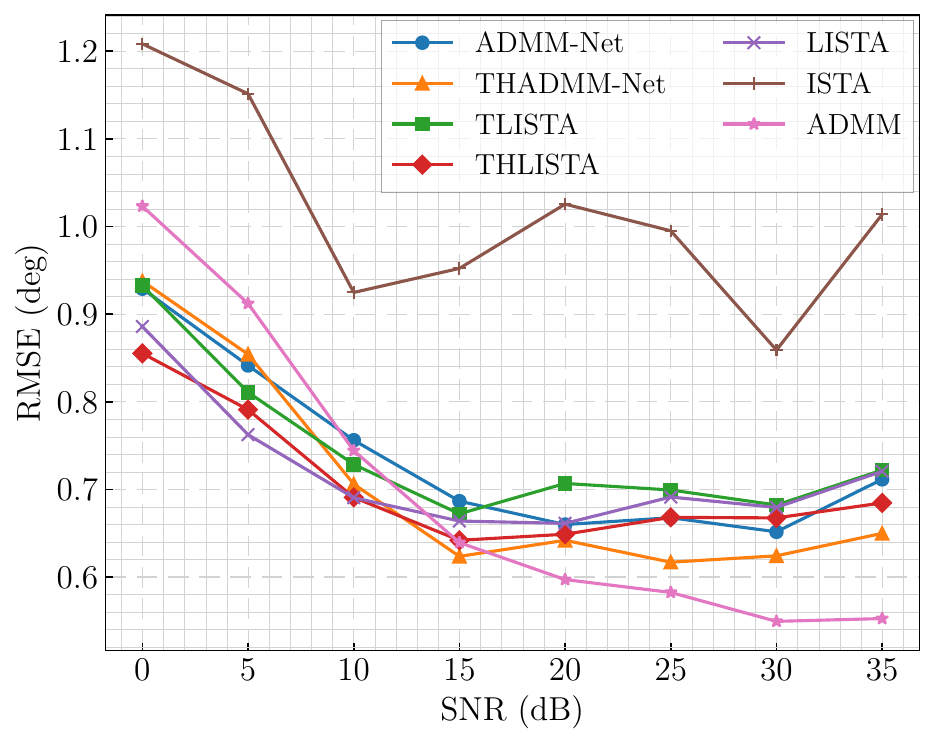}
			\caption{RMSE.}
			\label{rmse}
		\end{subfigure}		
	}
	\caption{Comparison of the (a) detection rate, (b) NMSE, and (c) RMSE performance of a 15-layer THADMM-Net against 30-layer LISTA, TLISTA, THLISTA, ADMM-Net as well as 50-iteration ADMM and 100-iteration ISTA.}
	\label{detect}
\end{figure*}
the estimated spectrum to the ground truth amplitude is higher than a given set threshold  $0<\delta_{2} \le 1$,
\begin{align}
	\dfrac{\mathbf{\hat{x}}_{pk}(j)}{|\mathbf{x}(q)|} \ge \delta_{2} , \qquad j \in \mathcal{J}_{q}.
\end{align}
If the set $\mathcal{J}_{q} $ is non-empty, the recovery of the ground truth target located at the bin frequency $q$ is considered successful. The procedure is repeated for the $K$ targets contained in a given ground truth vector $\mathbf{x}$. With the definition $\mathds{1}(\phi)= 1$ if $\phi \neq \emptyset$ and $0$ otherwise, the detection rate $P_{d}$ is computed as
\begin{align}
	P_{d} =\dfrac{\sum_{k=1}^{K}\mathds{1}(\mathcal{J}_{q(k)})}{K}.
\end{align}
For each SNR level, we then report the mean detection rate over a test set of 1000 measurement vectors.  For the $k$-th ground truth target located at the frequency bin $q(k)$, we first find the closest index $\tilde{j}(k)$ in $\mathcal{J}_{q(k)}$ to $q(k)$ 
\begin{align}
	\tilde{j}(k)=\argmin \limits_{j\in \mathcal{J}_{q(k)}} |j-q(k)|.
\end{align}
With the $n$-th bin of the angular grid defined as $\theta_{n}=\sin^{-1}(f_{n})/\gamma$, the mean squared angular error $E$ for a single test vector is then computed as
\begin{align}
	E=\dfrac{1}{N_{d}}\sum_{r=1}^{N_{d}}(\theta_{q(k_{r})}-\theta_{\tilde{j}(k_{r})})^{2}
\end{align}
where $N_{d}=\sum_{k=1}^{K}\mathds{1}(\mathcal{J}_{q(k)})$ is the number of successfully detected targets. For each SNR level, the mean of the angular error $E$ is averaged over the test vectors that have at least one successfully detected target and we report the square root of the result to yield the RMSE. In our experiments for the detection rate and RMSE, we fix $(\delta_{1}, \delta_{2})=(2, 0.4)$.  We additionally report the performance with respect to the NMSE.

% \begin{figure}[htb]
% 	\centering
% 	\includegraphics[height=3.2cm,width=4.6cm]{./plots/detection_rate_all.pdf}
% 	\caption{Detection rate across SNR of 15-layer THADMM-Net against 30-layerst TLISTA, THLISTA, ADMM-Net, and a 50-iteration ADMM along with 100 iteration of ISTA. }
% 	\label{fig:detection_rate}
% \end{figure}

\begin{figure}[!htb]
	\centering
	\begin{subfigure}{\linewidth}
		\centering
		\includegraphics[width=0.5\linewidth]{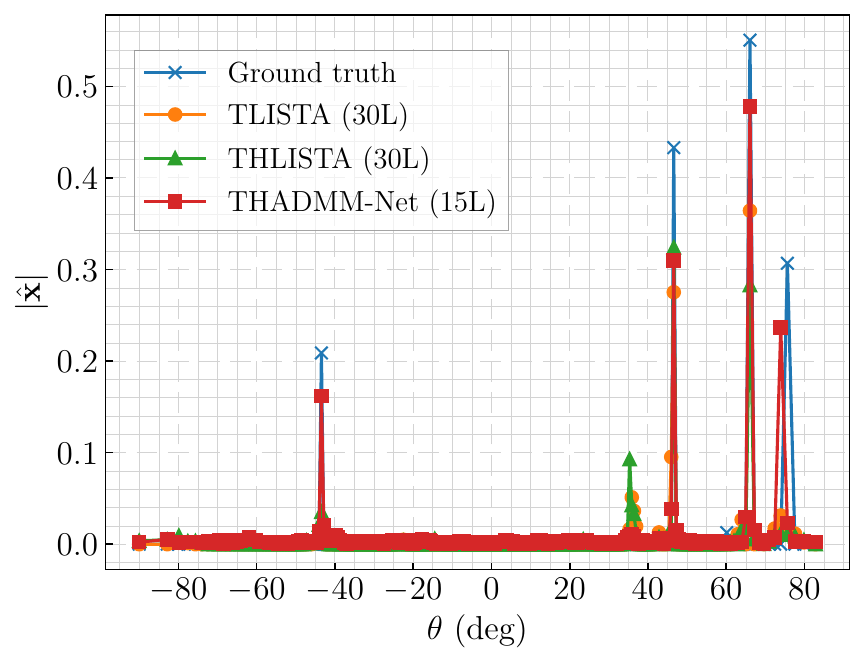}
		\caption{Spectrum sample at 15 dB.}
		\label{spec_15db}
	\end{subfigure}
	
	\begin{subfigure}{\linewidth}
		\centering
		\includegraphics[width=0.5\linewidth]{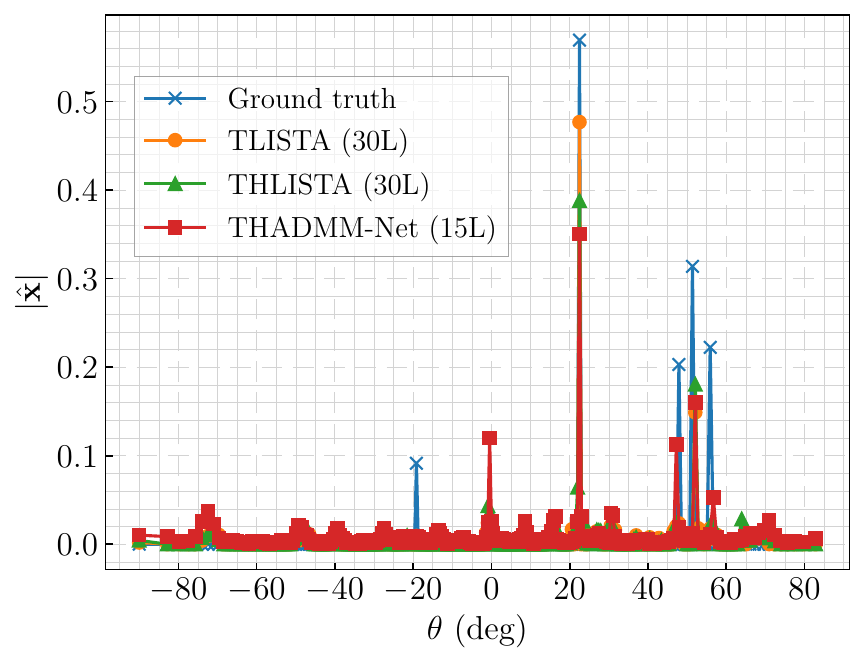}
		\caption{Spectrum sample at 5 dB.}
		\label{spec_5db}
	\end{subfigure}
	\caption{Spectrums of 15-layer THADMM-Net and 30-layer TLISTA, THLISTA for a 5 targets scenario under (a) 15 dB and (b) 5 dB SNR.}
	\label{losses}
\end{figure}

Fig. \ref{detection_rate} indicates the relative performance of the 15-layer THADMM-Net against the 30-layer TLISTA, THLISTA, and ADMM-Net, as well as a 50-iteration ADMM and 100-iteration ISTA. THADMM-Net outperforms TLISTA throughout the SNR range, with a detection rate of approximately 78\% against 60\% for SNRs above 15 dB. The same trend holds for the NMSE performance as can be seen from Fig. \ref{nmse} where THADMM-Net sits at about -4 dB in the high SNR region against -3 dB for TLISTA and THLISTA, we can also note that THADMM-Net shows a close performance relatively close to that of a 50-iteration of ADMM. We can additionally note from Fig. \ref{rmse} that THADMM-Net outperforms THLISTA with respect to the RMSE for SNRs above 10 dB, and shows a  close performance match with TLISTA for SNRs the 10 dB region. Fig. \ref{spec_15db} and Fig. \ref{spec_5db} show the reconstructed spectrums using THADMM-Net, TLISTA, and THLISTA for a 5-target scenario at 15 dB and 5 dB SNR, respectively. We can note THADMM-Net is able to recover more targets compared with TLISTA and THLISTA, which is consistent with the observed higher detection rate.

% \begin{figure}[h!]
% 	\centering
% 	\includegraphics[height=3.2cm,width=3.6cm]{./plots/spec_1_15.pdf}
% 	\caption{Spectrums of THADMM-Net, TLISTA, and THLISTA for a 5 targets scenario and 15 dB SNR.}
% 	\label{fig:spec_15db}
% \end{figure}

% \begin{figure}[h!]
% 	\centering
% 	\includegraphics[height=3.2cm,width=3.6cm]{./plots/spec_133_5dB.pdf}
% 	\caption{Spectrums of THADMM-Net, TLISTA, and THLISTA for a 5 targets scenario and 5 dB SNR. }
% 	\label{fig:spec_5db}
% \end{figure}

\section{Conclusion}

We proposed THADMM-Net, an efficiently-structured deep-unfolded version of ADMM that imposes a PSD as well as Toeplitz-Hermtian constraint on the learnable matrices. We showed that for a lower parameter count and depth, the resulting network outperforms TLISTA, THLISTA, LISTA and ADMM-Net with respect to the NMSE, the detection rate, and the RMSE across a range of SNRs. Future work includes extension of the proposed network to the 2D setting.
\newpage
\bibliographystyle{IEEEtran}
\bibliography{refs}

\end{document}